%% file: paper.tex
\input new-noteset

\documentclass[twocolumn,showpacs,preprintnumbers,amsmath,amssymb]{revtex4}

\usepackage{graphicx}
\usepackage{dcolumn}
\usepackage{bm}

\begin{document}

\preprint{APS/123-QED}

\title{\Large Shadowing-based reliability decay in softened $n$-body simulations}

\author{Wayne B. Hayes}
\email{wayne@cs.toronto.edu}
\affiliation{%
Department of Computer Science,
University of Toronto,
Toronto, Ontario, M5S 3G4
Canada
}%
\date{\today}

\begin{abstract}
A {\em shadow} of a numerical solution to a chaotic system
is an {\em exact} solution to the equations
of motion that remains close to the numerical solution for a long
time.  In a collisionless $n$-body system, we know that particle motion
is governed by the global potential rather than by
inter-particle interactions.  As a result, the trajectory of
each individual particle in the system is independently shadowable.
It is thus meaningful to measure the {\em number} of particles
that have shadowable trajectories as a function of time.
We find that the number of shadowable particles decays exponentially
with time as $e^{-\mu t}$, and that for $\eps\in [\sim 0.2,1]$ (in
units of the local mean inter-particle separation $\bar n$),
there is an explicit
relationship between the decay constant $\mu$, the timestep $h$ of the
leapfrog integrator, the softening $\eps$, and the number of particles
$N$ in the simulation.  Thus, given $N$ and $\eps$, it is possible to
pre-compute the timestep $h$ necessary to acheive a desired fraction of
shadowable particles after a given length of simulation time.
We demonstrate that a large fraction of particles remain shadowable over
$\sim$100 crossing times even if particles travel up to about 1/3 of the
softening length per timestep.  However, a sharp decrease in the number
of shadowable particles occurs if the timestep increases to allow particles
to travel further than 1/3 the softening length in one timestep, or if
the softening is decreased below $\sim 0.2\bar n$.
\end{abstract}


\maketitle

\paragraph*{\bf\label{sec:intro} Introduction.}

Numerical simulation of the softened gravitational $n$-body problem is
used to gain insight into the formation, evolution and structure of
gravitational systems ranging from galaxies and clusters of galaxies
to large-scale stucture of the Universe
\cite{ClarkeWest97,Bertschinger98}.  Since such simulations have been
used to invalidate theories \cite{Bertschinger98}, establishing
their trustworthiness is critical.
These simulations have several sources of error, including:
the use of several orders of magnitude too few particles, or {\it
discreteness noise};
the use of approximate force-computation methods (the latter two errors
are compared in \cite{HernquistHutMakino93});
the use of a softened potential;
the use of finite-timestep numerical integration to evolve
the system of ordinary differential equations;
and machine roundoff error.
These errors are aggravated by the fact that gravitational $n$-body
systems are chaotic and display {\it sensitive dependence on initial
conditions}: two solutions with nearby initial conditions
diverge exponentially away from each other on about a crossing
timescale \cite{GoodmanHeggieHut93},
so that {\em any}
error results in the numerical trajectory diverging exponentially away
from the exact solution with the same initial conditions.
The phenomenon has been described (\eg \cite{GoodmanHeggieHut93})
as the ``exponential magnification of small errors'', implying the
possibility that trajectories of such simulations are the result of
nothing but magnified noise.

Fortunately, the purpose of a softened $n$-body simulation is not to
follow the evolution of a particular choice of initial conditions, but
instead to {\em sample} the evolution of large systems whose initial
conditions are drawn from a random distribution.  As such, we would
likely be more than satisfied if our numerical solution closely
followed the evolution of a nearby set of initial conditions.  The
study of {\it shadowing} provides just such a property: a {\it shadow}
of a numerical, or {\it noisy}, solution is an exact solution whose
initial conditions and subsequent evolution remain nearby, in phase-space, to
the numerical solution.  Thus, a numerical solution that has a
shadow is essentially an experimental observation of an exact
trajectory of the mathematical system being modelled.  Although this
observation does not alleviate errors introduced between the physical
system and the mathematical model (such as discreteness noise and force
softening), it {\em does} say that the numerical simulation is
faithfully solving the mathematical model.  In \cite{QT}, it was shown
that a single particle moving amongst 99 fixed particles is shadowable
for several tens of crossing times, and that {\it glitches} (the point
beyond which a shadow cannot be found) tend to occur near close
encounters.  In \cite{HayesM1}, we demonstrated that if $M>1$ particles
move in a softened system with $100-M$ fixed particles, then very few
particles encounter glitches within the first few tens of crossing
times.  However, both of these studies used highly accurate integrators
to generate the ``noisy'' trajectories.  Although high accuracy is
commonly used for simulations of unsoftened systems, softened
simulations most often use the 2nd-order symplectic and time-symmetric
{\it leapfrog} integrator.

\begin{figure}
\includegraphics[width=8cm,height=3cm]{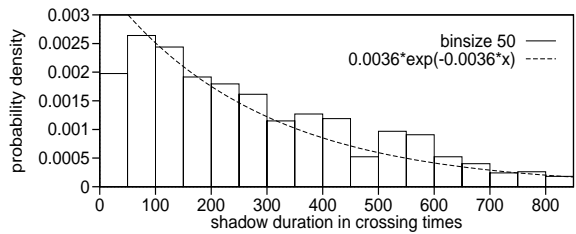}
\caption{Histogram of shadow durations of 1000 systems in which 1 particle
moves in the potential of 99 fixed particles with softening $\eps=0.054$ or
1/4 of the mean inter-particle spacing.
Noisy trajectories were generated using a leapfrog integrator
with timestep $h=0.011=\eps/5$.
After an initial transient, the distribution fits
an exponential curve with a mean shadow duration of 280 crossing times,
indicating that the moving particle encounters glitches as a
Poisson process with a glitch probability of 0.36\% per crossing time.}
\label{fig:M1-k5-distribution}
\end{figure}

\begin{figure}
\includegraphics[width=8cm,height=4cm]{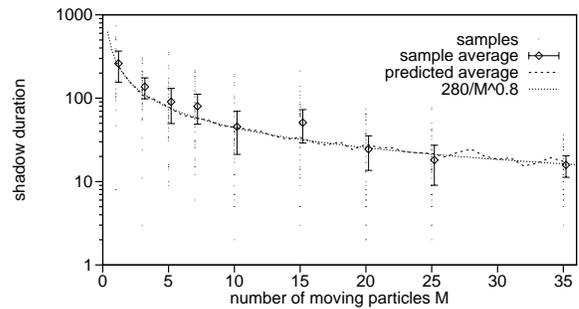}
\caption{How the average shadow duration scales as the number of
moving particles $M$ is increased.
Each system is identical to that described in Figure
\ref{fig:M1-k5-distribution}, except now $M$ takes on the values
$1,3,5,7,10,15,20,25,35$.
The dots represent sample shadow durations, 40 samples for each $M$,
while the ``sample average'' is plotted with sample error bars of full
width $1\sigma$.
The ``predicted average'' is artificially constructed
for each $M=1,\ldots,35$ by superimposing $M$ samples chosen at random
from Figure \ref{fig:M1-k5-distribution} and taking the minimum shadow
duration of those samples.  This demonstrates that the average shadow
duration of an $M$-moving-particle system is well-predicted using
$M$ single-moving-particle systems, and suggests that particles encounter
glitches independently of one another; $280/M^{0.8}$ is plotted for
comparison, although the physical significance of 0.8 is unclear.}
\label{fig:k5-vs-M}
\end{figure}

\begin{figure}
\includegraphics[width=8cm,height=4cm]{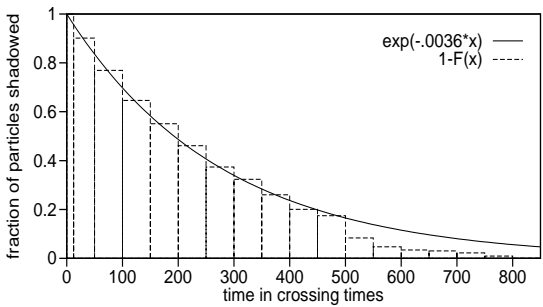}
\caption{The estimated fraction of particles that would be shadowed as
a function of time, in a
system similar to that described in Figure \ref{fig:M1-k5-distribution}.
except with {\em all} particles moving.
This is computed simply by
taking the cumulative distribution function $F(t)$ of Figure
\ref{fig:M1-k5-distribution}, where $F(t)$ represents the fraction
of particles that have undergone a glitch, and then plotting $1-F(t)$.}
\label{fig:k5-rhist}
\end{figure}

\begin{figure}
\includegraphics[width=8cm,height=5cm]{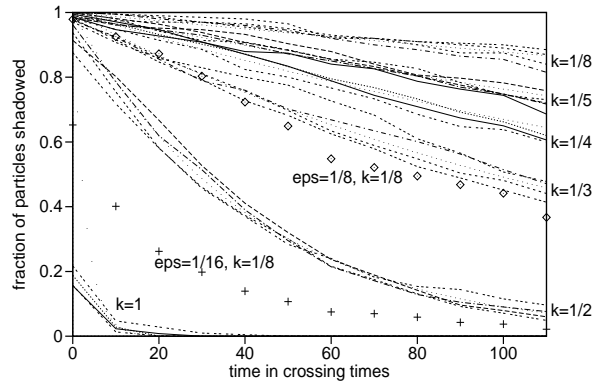}
\caption{Curves demonstrating that
(i) Equation (\ref{eq:h=eps2N3}) preserves simulation reliability
from a shadowing standpoint,
(ii) reliability increases as $h$ decreases, and
(iii) the scaling breaks down if $\eps$ is too small.
Each line represents a curve similar to that
in Figure \ref{fig:k5-rhist}, derived from a 200-sample set of
simulations similar to Figure \ref{fig:M1-k5-distribution} but with
different $N$ and $\eps$.
There are 6 clusters of 4 lines each.
The 4 lines in each cluster come from 4 sets of simulations using all
4 combinations of $N=100,1000$ and $\eps=\{\frac{1}{2},\frac{1}{4}\}N^{-1/3}$.
Each system was integrated using leapfrog with timestep $h$ scaled
according to Equation (\ref{eq:h=eps2N3}).
Each cluster represents a particular choice of constant $k$ in the scaling,
namely, $h^2 = k^2 \eps^2 (\frac{N}{100})^{1/3}$, for the displayed values
of $k$.  Decreasing $k$ increases accuracy, giving longer shadows.
Embedded in each cluster is a dotted line representing $e^{-\mu_k t}$,
discussed later.
Finally, the diamond and '+' curves have
$\eps=\{\frac{1}{8},\frac{1}{16}\}N^{-1/3}$, respectively, and
demonstrate that shadows are much shorter if the softening is too small,
even though they use the smallest timestep factor $k=\frac{1}{8}$.
}
\label{fig:all-rhist-notitles}
\end{figure}

\begin{figure}
\includegraphics[width=8cm,height=5cm]{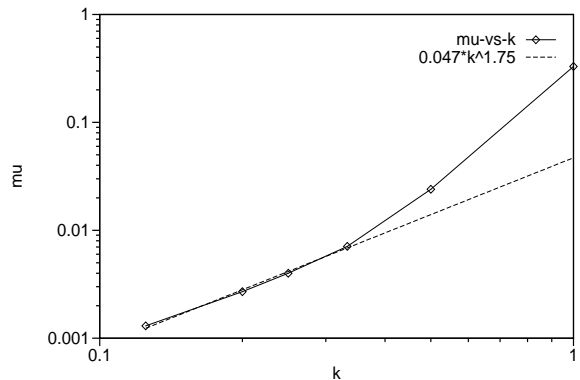}
\caption{For $k=1, \frac{1}{2}, \frac{1}{3}, \frac{1}{4},
\frac{1}{5}, \frac{1}{8}$ and $\eps\ge\frac{1}{4}N^{-1/3}$ of
Figure \ref{fig:all-rhist-notitles}, the values of $\mu_k$ were fit by eyeball,
and are, respectively, $\mu=0.33, 0.024, 0.0071, 0.0040, 0.0027, 0.0013$.
These are plotted, and for $k\lesssim \frac{1}{3}$, fit the curve
$\mu=(0.047\pm 0.015) k^{1.75\pm 0.25}$.}
\label{fig:mu-vs-k}
\end{figure}

In this paper, we use the leapfrog integrator to generate noisy
trajectories of systems that have $M$ particles moving and interacting
in a softened potential amongst a background of $N-M$ fixed particles
\cite{QT,HayesMSc,HayesM1}.  We use normalized units
\cite{HeggieMathieuNoXRef} in which each particle has mass $1/N$, and
the system diameter, crossing time and average velocity all have order
unity.  We then lead the reader through the following observations.
First, we observe that glitches in the trajectory of a single particle
occur as a Poisson process (Figure \ref{fig:M1-k5-distribution}).  Next,
we demonstrate that as $M$ increases, shadow durations scale roughly as
$1/M^{0.8}$.  (The physical significance of 0.8 is unclear and
may be dependent upon other parameters such as $N$, the softening
$\eps$, and the timestep $h$.)  More importantly, this scaling can be
experimentally reproduced by superimposing trajectories of $M$
single-moving-particle systems and
taking the {\em shortest} shadow of those $M$ systems.  In other words,
particles appear to encounter glitches independently of one another
(Figure \ref{fig:k5-vs-M}).
Now, the glitching and subsequent errant behaviour of just one particle
(the first to undergo a glitch) in a large simulation is unlikely to
have a large effect on the reliability of that simulation; in fact,
as long as only a small fraction of particles have
glitched, then the reliability of the simulation probably remains high.
Then, assuming that particles encounter glitches independently of
one another, we can use the distribution of shadow durations of $M=1$
systems to predict
the {\em fraction of shadowable particles} as a function of time.
We find that this fraction is a decaying exponential with some exponent $\mu$
(Figures \ref{fig:k5-rhist} and \ref{fig:all-rhist-notitles}).
Finally, we demonstrate an explicit relationship between $\mu$,
$N$, $\eps$, and $h$ that holds as long as $\eps$ is in the range
$\sim[0.2,1]$ times the mean inter-particle separation
and $h\lesssim\eps/3$ (Figure \ref{fig:mu-vs-k}).
This means that given $N$, $\eps$, and the expected
duration of the simulation, one can {\em pre-compute} the timestep $h$
necessary to have a desired fraction of shadowable particles remain at
the end of the simulation.

\paragraph*{\bf Results.}
Figure \ref{fig:M1-k5-distribution} introduces a histogram of shadow
durations for 1000 softened systems with $N=100,M=1$.  After
an initial transient (explained in the discussion of
\cite[Figure 3]{HayesM1}), the distribution fits an exponential curve,
suggesting that glitches occur as a Poisson process.

Figure \ref{fig:k5-vs-M} introduces how the average shadow duration
scales as the number of moving particles is increased.  For various
values of $M$, we perform 40 experiments in which $M$ particles
move and interact amongst $100-M$ fixed pariticles, and plot the
mean and standard deviation of the shadow durations.
We make the following observations:
(1) a glitch in the local 6-dimensional phase-space trajectory of any
one particle will cause a glitch in the full $6M$-dimensional
phase-space trajectory of the $M$-moving particle system.
(2) in a large collisionless system,
the gravitational potential is governed more by the global potential
than by inter-particle interactions \cite{BinneyTremaine},
and so it is reasonable to expect
that particles encounter glitches independently of each other.
So, perhaps the mean shadow duration of an $M$-moving-particle
system can be predicted by the mean shadow duration of a system with
$M$ completely uncoupled 1-moving-particle systems \cite{HayesM1}.
We test this hypothesis with the ``predicted average'' of Figure
\ref{fig:k5-vs-M} by taking $M$ samples at random from Figure
\ref{fig:M1-k5-distribution} and taking the {\em shortest} shadow
duration.  We see that the predicted curve is well within the error
bars of the ``real'' $M$-moving-particle system.  Formally, this
suggests that {\em the average duration before the occurance of the
first glitch in the system is statistically independent of whether
particles interact or not.}

Once one particle in the system
encounters a glitch, its trajectory after that point is incorrect,
and it will presumably start to ``infect'' the motion of other
particles.  However, by observation (2) above, we can hope that
in a large collisionless system, one errant particle will, for
a time, have negligible effect on the trajectories of other particles.
In fact, we can guess that, for a time, any small fraction of errant
particles will have little effect on the global behaviour of system.
The goal would then be to minimize, at reasonable cost, the number
of particles that have glitched by the end of a given simulation.

We now take as a working assumption that particles encounter glitches
independently of one another, and that the first small fraction of particles
that encounter glitches have a negligible effect on the others.
That is, we re-interpret Figure \ref{fig:M1-k5-distribution} to represent
the distribution of shadow durations for {\em all} the individual particles
in a {\em single} many-moving-particle system.
Of course, the figure is likely only to
be valid for a duration much shorter than 800 crossing times, as the
earlier glitched particles ``infect'' the motion of the remainder, but
let us assume it is a reasonable approximation for some shorter period.
This allows us, as a first approximation, to estimate the fraction
of glitched particles in a real simulation at a given time by computing
the fraction of one-moving-particle systems that have glitched by that time.
Figure \ref{fig:k5-rhist} plots the opposite---the fraction of non-glitched
(\ie shadowed) particles as a function of time---derived from
Figure \ref{fig:M1-k5-distribution} by taking its cumulative distribution
function $F(t)$ and ``flipping'' it to $1-F(t)$.
As expected from the Poisson process in Figure \ref{fig:M1-k5-distribution},
the fraction of shadowable particles decays exponentially with a rate
corresponding to a 0.36\% glitch probability per particle per
crossing time.  

Figure \ref{fig:k5-rhist} is interesting, but is of little use because
it does not tell us how the shape of this curve varies with the total
number of particles $N$, the softening $\eps$, or the timestep $h$.
However, for reasons we will discuss later, we have found that if the
timestep $h$ is scaled as
\begin{equation}
h^2 \propto \eps^2 N^{1/3},
\label{eq:h=eps2N3}
\end{equation}
then each of Figures 1, 2, and 3 are preserved if $\eps$ is not too small.
That is, if $N$ and
$\eps$ are changed in a given simulation but the initial conditions are
drawn from the same distribution, then {\em using Equation
(\ref{eq:h=eps2N3}) to scale the timestep will preserve the same degree
of simulation reliability from the standpoint of shadowing}.
Intuitively, the scaling of $h\propto\eps$ is not surprising and is a
commonly used timestep criterion.
The $N^{1/3}$ is
more surprising, telling us we can {\em increase} the timestep as $N$
increases at a fixed softening; intuitively, this is because the
gravitational potential becomes more smooth with increasing $N$.

To demonstrate this scaling, we have performed many experiments with
various values of $h$, $N$, and $\eps$.  Figure \ref{fig:all-rhist-notitles}
summarizes the results.
Each line represents the fraction of shadowable
particles as a function of time for some $(N,\eps)$ pair.  Each closely
clustered set of 4 lines represents sets of runs with various $(N,\eps)$
and the the timestep $h$ scaled using Equation \ref{eq:h=eps2N3} to
give the same ``shadowing reliability''.
Finally, decreasing the timestep (via decreasing $k$)
increases reliability by decreasing the decay rate of the fraction of
shadowed particles.
We also performed similar experiments with all combinations
of parameters $N=10^2,10^3,10^4$ and
$\eps=\{2, 1, \frac{1}{2}, \frac{1}{4}, \frac{1}{8}, \frac{1}{16}\}N^{-1/3}$.
The scaling with $N$ works well up to $N=10^4$, and presumably beyond.
All curves are very similar for $\eps\ge\frac{1}{4} N^{-1/3}$.
However, as seen in the figure, shadows are significantly shorter
for $\eps=\{\frac{1}{8},\frac{1}{16}\}N^{-1/3}$, even if the smallest
timestep is used.
This is consistent with the observations of \cite{QT},
and may be related to unphysical results obtained with a too-small
softening \cite{SplinterMelottShandarinSuto98}.

Finally, we come to the crux.  Armed with the scaling Equation
(\ref{eq:h=eps2N3}) and the knowledge that the fraction of shadowable
particles decays exponentially as $e^{-\mu_k t}$, we would like to
find a relationship between the timestep proportionality constant
$k$ of Figure \ref{fig:all-rhist-notitles}, and $\mu_k$.  An eyeball
fit of exponential curves to each of the clusters in Figure
\ref{fig:all-rhist-notitles} gives values of the decay constant $\mu_k$ for
each cluster.  These values, along with the curve $\mu_k$ \vs $k$, is plotted
in Figure \ref{fig:mu-vs-k}.  As can be seen, there is evidence that
for $k\lesssim\frac{1}{3}$, the curve settles to a power law of
approximately
\begin{equation}
\mu_k = (0.047\pm 0.015) k^{1.75\pm 0.25}.
\label{eq:mu-vs-k}
\end{equation}
However, the shape of the curve is also consistent with a slowly
decreasing slope as $k$ decreases.


\paragraph*{\bf Discussion.}

The relations described above offer an {\it a priori} algorithm for
choosing a timestep for a softened $N$-body simulation, \viz: given
$N$, $\eps$, the expected simulation duration $T$ in crossing times,
and a desired fraction $F$ of shadowable particles remaining at time $T$,
solve for $\mu$ in $F=e^{-\mu T}$, and then solve for $k$ using
Equation (\ref{eq:mu-vs-k}).  Of course, these relationships will need
to be scaled to appropriate units for the simulation.
The job should be easiest for a simulation of one galaxy;
for simulations of clusters of galaxies or a cosmological simulation,
we would scale to the smallest sub-systems we expect to accurately
integrate.
We are unsure of the effects of dynamically changing the softening based
on the local mean particle density, but suspect that some reasonable
interpretation
may be possible whereby a softening and timestep (modulo the
discussion of the next paragraph) are chosen based upon local mean particle
density.

The fact that constant-timestep leapfrog is symplectic is probably
significant to these results.
We experimented briefly with a dynamically changing timestep, but
found that shadows were virtually destroyed if the timestep changes
``too often''.  However, we found that if the timestep was decreased
as a particle entered a high-density region, but never increased the
timestep again, these results were preserved.  This may be a reasonable
choice for simulations of clusters if most particles that enter a
high-density region remain there for the remainder of the simulation.
Alternatively, perhaps a particle's timestep could be re-increased only
after the dynamic timestep criterion says the particle's timestep should
be {\em significatly} increased, say by an order of magnitude.  This will
ensure the particle has left the high-density region far behind, and
preserve the internal reliability of high-density regions.

More detailed arguments deriving Equation (\ref{eq:h=eps2N3})
show that the {\it forward global error}
of a softened $N$-body simulation scales as $h^2\eps^{-2}N^{-1/3}$.  The
$h^2$ scaling is due to leapfrog being a globally 2nd-order integrator;
the $N^{-1/3}$ scaling has been seen before \cite{GoodmanHeggieHut93},
and the $\eps^{-2}$ is new.  Thus, Equation (\ref{eq:h=eps2N3}) simply
holds the forward global error constant.  Finally, a shadowing concept
known as {\it brittleness} \cite{DawsonGrebogiSauerYorke94}
relates the forward global error to the distance between a shadow
and the corresponding points on the numerical trajectory,
although to our knowledge this paper
is the first to demonstrate a relationship between the the forward global
error and the shadowing distribution.

Since the scaling of $h$ with $N^{1/3}$ is so weak, and we
usually increase $N$ in order to increase reliability, it certainly
will not hurt to ignore the scaling with $N$ and simply scale
$h\propto\eps$.  The remaing questions are: what value of $\eps$ to use,
and what fraction of $\eps$ should a particle be allowed to travel
in one timestep?  The first question is answered by the fact that
shadows do not appear to exist for very long if
$\eps\lesssim\frac{1}{8}N^{-1/3}$; concerning the second question,
we believe that Figures \ref{fig:all-rhist-notitles} and \ref{fig:mu-vs-k}
suggest that the particle
should be allowed to travel {\em at most} $\frac{1}{3}\eps$ in one timestep,
and possibly much less, if a shadow duration of $100$ crossing
times encompassing most particles is desired.

{\bf Acknowledgments.} We thank A. Melott for suggestions on the presentation
of these results.
\bibliography{paper}
\end{document}

%% file: new-noteset.tex
\newcommand{\eps}{{\varepsilon}}

\def\eg		{{\it eg.,~}}

\def\ie		{{\it i.e.,~}}

\def\vs		{{\it vs.~}}
\def\viz	{{\it viz.~}}

